\documentclass[a4paper,11pt]{article}
\usepackage{harvard}

\usepackage[dvips]{graphicx}
\usepackage{amssymb}
\usepackage{amsmath}
\usepackage{amsthm}

\usepackage{url}

\newcommand \be{\begin{equation}}
\newcommand \bea{\begin{eqnarray}}
\newcommand \ee{\end{equation}}
\newcommand \eea{\end{eqnarray}}

\newcommand{\E}{{\rm E}}

\renewcommand{\(}{\left(}
\renewcommand{\)}{\right)}
\renewcommand{\[}{\left[}
\renewcommand{\]}{\right]}
\renewcommand{\epsilon}{\varepsilon}

\theoremstyle{plain}

\theoremstyle{definition}

\begin{document}

\title{Gibrat's law for cities: \\
uniformly most powerful unbiased test \\of the Pareto against the lognormal
}
\thispagestyle{empty}

\author{Y. Malevergne$^{1,2,4}$, V. Pisarenko$^{3}$ and D. Sornette$^{4,5}$\\
$^1$ Universit\'{e} de Lyon -- Universit\'{e} de Saint-Etienne -- Coactis E.A. 4161, France\\
$^2$ EMLYON Business School -- Cefra, France\\
$^3$ International Institute of Earthquake Prediction Theory and Mathematical Geophysics\\
Russian Academy of Science, Moscow, Russia\\
$^4$ ETH Zurich -- Department of Management, Technology and Economics, Switzerland\\
$^5$ Swiss Finance Institute, Switzerland\\
e-mails: malevergne@em-lyon.com, pisarenko@yasenevo.ru and dsornette@ethz.ch
}

\date{}
\maketitle

\begin{abstract}
We provide definitive results to close the debate between Eeckhout (2004, 2009) and  \citeasnoun{Levy2007}
on the validity of Zipf's law, which is the special Pareto law with tail exponent $1$,
to describe the tail of the distribution of U.S. city sizes. 
Because the origin of the disagreement between Eeckhout and Levy
stems from the limited power of their tests, we perform the {\em uniformly most powerful unbiased 
test} for the null hypothesis of the Pareto distribution against the lognormal.
The $p$-value and Hill's estimator as a function of city size lower threshold confirm
indubitably that the size distribution of the 1000 largest cities or so,
which include more than half of the total U.S. population, is Pareto, but we
rule out that the tail exponent, estimated to be $1.4 \pm 0.1$, is equal to $1$.
For larger ranks, the $p$-value becomes very small and Hill's estimator decays
systematically with decreasing ranks, qualifying the lognormal distribution as the better model for the set of smaller cities. 
These two results reconcile the opposite views of  \citeasnoun{Eeckhout2004} and  \citeasnoun{Levy2007}.
We explain how Gibrat's law of proportional growth underpins both the Pareto 
and lognormal distributions and stress the key ingredient at the origin of their difference in
standard stochastic growth models of cities  \cite{Gabaix99,Eeckhout2004}.

\end{abstract}

\vspace{2cm}

\noindent
{\bf JEL classification:} D30, D51, J61, R12.
\vspace{0.5cm}

\noindent
{\bf Keywords:} City sizes, Gibrat's law, Zipf's law.

\clearpage

Based upon the U.S. Census 2000 data, \citeasnoun{Eeckhout2004} reports that the whole size distribution of cities is lognormal rather than Pareto. This conclusion is obtained by using the Lilliefors test (L-test) \cite{Lilliefors67,Stephens74}  for normal distributions with empirical mean ${\hat \mu} = 7.28$ and standard deviation $\hat \sigma = 1.25$. This empirical conclusion is consistent with Gibrat's law of proportionate effect and is rationalized by an equilibrium theory of local externalities in which the driving force is a random productivity process of local economies and the perfect mobility of workers.

\citeasnoun{Levy2007} argues that the top 0.6\% of the largest cities of the U.S. Census 2000 data sample, which accounts for more than 23\% of the population, dramatically departs from the lognormal distribution and is more in agreement with
a power law (Pareto) distribution. The bulk of the distribution actually follows a lognormal but, due to the departure in the upper tail, a $\chi^2$-test unequivocally rejects the null of a lognormal for cities whose log-size is larger than $\hat \mu + 3 \hat \sigma = 12.53$. The non-rejection of the lognormal by the L-test used by \citeasnoun{Eeckhout2004} is ascribed to the fact that the relative number of cities in the upper tail is very small (only 0.6\% of the sample), and the L-test is dominated by the center of the distribution rather than by its tail, where the interesting action occurs.

In reply, \citeasnoun{Eeckhout2008} provides the 95\%-confidence bands of the lognormal estimates based upon the L-test and shows that the tail of the sample distribution of log-size is well within the confidence bands. In addition, Eeckhout asserts that ``both [Pareto and lognormal] distributions are regularly varying, i.e. they are heavy tailed, and their tails have similar properties. [...] It is natural that the upper tail of city sizes can be fit to a Pareto distribution''. Therefore ``[g]iven that the tail of a lognormal is indistinguishable from the Pareto under certain circumstances, the researcher who is interested in the tail properties of a size distribution can choose which one to use.''  

In the first part of this comment, we summarize the properties that make often difficult the task
of distinguishing between the Pareto and the lognormal distributions. While the 
Pareto and the lognormal distributions have indeed distinct
asymptotic tails -- in contrast with the Pareto, the lognormal {\em is not} regularly varying but rapidly varying --
the lognormal can easily be mistaken for a Pareto over a range which
can cover several decades as soon as its standard deviation is sufficiently large (a few units is sufficient). Furthermore, 
both distributions may be generated by Gibrat's law of proportional growth, with some
additional apparently innocuous but actually profound twist(s) for the Pareto.
In a second part, using exactly the same data set, 
we find that the origin of the disagreement between Eeckhout and Levy
stems from the limited power of their tests. Using the {\em uniformly most powerful unbiased 
test} for the null hypothesis of a Pareto distribution against the lognormal, we confirm and extend Levy's result,
by showing that the Pareto model holds for the 1000 largest cities or so, i.e. for more than 50\% of the total population. 
Zipf's law, corresponding to Pareto with exponent $1$, is found incompatible with the data
at the 90\% confidence level. The Pareto index for the uppermost tail (about 1000 largest cities) is approximately $1.4$.

\section{Why the Pareto and the Lognormal distributions are difficult to distinguish}

\subsection{Structural similarities and differences}

In order to justify that Levy's results are compatible with his own, \citeasnoun{Eeckhout2008} asserts that both the Pareto distribution and the lognormal distribution are regularly varying, which makes their tail indistinguishable. We recall that a positive function $f(x)$ is regularly varying at infinity if there exists a finite real number $\alpha$ such that \cite{BGT87}
\be
\lim_{x\to \infty} \frac{f(t\cdot x)}{f(x)} = t^\alpha, \quad \forall t>0.
\ee
Pareto distributions are regularly varying. However, it is not the case for lognormal distributions. Indeed, the lognormal density reads
\be
f(x) = \frac{1}{\sqrt{2 \pi} \sigma} \cdot \frac{1}{x}\, e^{- \frac{(\ln x - \mu)^2}{2 \sigma^2}}~,
\ee
so that
\be
\lim_{x\to \infty} \frac{f(t\cdot x)}{f(x)} = \lim_{x\to \infty} \frac{1}{t} \,  e^{-\frac{(\ln t)^2}{2 \sigma^2}} e^{-\ln t \cdot \frac{\ln x - \mu}{\sigma^2}} = 
\begin{cases}
0, & t >1,\\
1, & t=1,\\
\infty, & t<1.
\end{cases}  
\ee
This limit behavior characterizes a rapidly decreasing function at infinity. Therefore, Pareto and lognormal distributions exhibit {\it qualitatively} different behaviors in their upper tails. The lognormal density goes to zero, in the upper tail, faster than any Pareto density. In this respect, they cannot be mistaken into one another, provided that one has enough data to sample the tail.

However, writing the lognormal density as follows
\be
f(x) = \frac{1}{\sqrt{2 \pi} \sigma} \cdot \frac{1}{x}\, e^{- \frac{(\ln x - \mu)^2}{2 \sigma^2}} = \frac{1}{\sqrt{2 \pi} \sigma} e^{- \frac{\mu^2}{2 \sigma^2}} \cdot x^{-1 + \frac{\mu}{\sigma^2} - \frac{\ln x}{2 \sigma^2}}~, 
\label{thertbww}
\ee
we observe that the lognormal distribution is superficially like a Pareto distribution with a slowly increasing
effective exponent
\be
\alpha(x) = {1 \over 2 \sigma^2} \ln \left( {x \over e^{2 \mu}}\right)~.
\label{tbwrtbqga}
\ee
Expression (\ref{tbwrtbqga}) allows us to make two points. First, as stated above, it shows that the lognormal distribution decays at infinity faster than any Pareto distribution, since the apparent exponent $\alpha(x)$ diverges with $x$.
Second, if $\sigma^2$ is large enough, the apparent exponent $\alpha(x)$ varies
so slowly so as to give the impression of constancy over several decades in $x$. Quantitatively,
in the range $X \leq x \leq \lambda X$, the apparent exponent varies from 
$\alpha(X)$ to $\alpha(X) + {1 \over 2 \sigma^2} \ln \lambda$. For instance, for $\sigma =3.4$,
the apparent exponent varies by no more than $0.3$ over three decades ($\lambda=1000$).

However, with the smaller estimate $\hat \sigma =1.25$ provided by \citeasnoun{Eeckhout2004}
for the U.S. Census 2000 data,
the apparent exponent varies by $1.5$ units over just two decades. This is an indication
that a powerful test, as implemented in the next section, should be able to distinguish
the two hypotheses over a range of two to three decades corresponding to the 
tail regime suggested by \citeasnoun{Levy2007}.

\subsection{Generating process}

Gibrat's law of proportional growth is often taken as a key starting point to understand the origin
of the distribution of city sizes (see the recent review by \citeasnoun{SaichevetalZipf09} and references therein).
\citeasnoun{Eeckhout2008} also stressed that Gibrat's law remains the corner stone for building economic models of population dynamics.  Considered as the unique ingredient, Gibrat's law predicts that the distribution of city sizes should tend to a lognormal distribution, but as a more and more degenerate one as time increases. Indeed, Gibrat's law leads
to model the growth of a given city as following a random walk in its log-size, which therefore never admits a steady state distribution. 

The equation of city growth embodying GibratÍs law is
\begin{equation}
S_{i,t}= a_{i,t}  \cdot S_{i,t-1}~,           
\label{thj2tk21t}
\end{equation}
where $S_{i,t}$ is the the size of city $i$ at time $t$ and $a_{i,t}$ is the random positive growth factor. 
Taking the logarithm of (\ref{thj2tk21t}) and iterating yields
\begin{equation}
\ln S_{i,t}= \ln S_{i,t-1} + \eta_{i,t} = \ln S_{i,0} + \eta_{i,1} + \eta_{i,2} + ...+ \eta_{i,t} ~,           
\label{thj2th2t24k21t}
\end{equation}
where $\eta_{i,t} \equiv \ln a_{i,t}$.
Assuming (for a time) that terms $\eta_{i,t}$ are iid 
random variables with expectation $A$ and standard deviation $B$, the Central Limit Theorem of Probability Theory
gives
\begin{equation}
\ln S_{i,t} \simeq  t \cdot A + t^{1/2} B \cdot \xi~,                                                        
\label{hjtkkkwfd}
\end{equation}
where $\xi$ is a standard Gaussian random variable $N(0,1)$. Of course, the stationarity of the $\eta_{i,t}$'s should be verified by an appropriate analysis. Assuming in addition that the stochastic growth process 
for a typical city as a function of time is equivalent to sampling the growth of many cities at a given instant,
i.e., that a strong form of ergodicity holds, expression (\ref{hjtkkkwfd}) ensures that the
the distribution of city sizes is lognormal, i.e.,  the variable
${\ln S_{i,t} - t \cdot A \over t^{1/2} B}$ is $N(0,1)$.

As recalled for instance by \citeasnoun{Gabaix99}, an apparently minor modification leads to a bona fide steady state and, therefore, to a stationary distribution of city sizes. This modification, which can take many forms \cite{sor98}, consists
in preventing the small cities from becoming too small. The corresponding generic equation of motion for city sizes embodying this idea together with Gibrat's law is \cite{Gabaix99}
\be
\label{eq:motion}
S_{i,t} = a_{i,t} \cdot S_{i,t-1} + \epsilon_{i,t} ~,
\ee
where the terms $\epsilon_{i,t} > 0$ prevent the accumulation of a large number of cities with vanishingly small sizes. In absence of $\epsilon_{i,t}$, expression (\ref{eq:motion}) is nothing but the random walk in log-size leading to the lognormal distribution obtained from (\ref{hjtkkkwfd}). Because the process (\ref{eq:motion}) with non-zero $\epsilon_{i,t}$ leads to a stationary distribution,\footnote{The condition for stationarity is $\E \[ \ln a_{i,t} \] <0$.} if we assume ergodicity, then the distribution of an ensemble of cities is the same as that of the set of realizations $\{S_{i,t}\}$ for a fixed city $i$ as a function of $t$ for large times.

The presence of the ``minor modification'' $\epsilon_{i,t}>0$ ensures that the size distribution of cities switches from a lognormal to a Pareto, even if it is arbitrarily small, as long as it is non-zero \cite{Kesten73}. 
The tail index $\alpha$ of the Pareto distribution is the solution to $\E \[(a_{i,t})^\alpha\] =1$. 
\citeasnoun{Gabaix99} argued for the validity of the constraint $\E \[a_{i,t}\] =1$, which then leads to 
Zipf's law: $\alpha =1$.  \citeasnoun{SaichevetalZipf09} shows that Zipf's law is more realistically the
result of Gibrat's law together with a condition balancing the birth rate, random growth and possible 
death rate of cities\footnote{In the case of cities, death means falling below a moving threshold for qualifying
as a city.}. 

The intuition behind the transformation of the lognormal into the Pareto distribution, upon the introduction
of the apparently minor additive term $\epsilon_{i,t} > 0$ is the following.  Because of the
stationarity condition $\E \[ \ln a_{i,t} \] <0$, in the absence of $\epsilon_{i,t}$, the process $S_{i,t}$
tends to shrink stochastically towards zero, while exhibiting a more and more degenerate
lognormal distribution. During this phase, a few excursions of exponentially large sizes
associated with transient occurrences of the growth factor $a_{i,t}$ larger than $1$
can occur with exponentially small probability. The term  $\epsilon_{i,t}$
allows the process to repeatedly exhibit the exponentially rare
exponentially large excursions. The combination of these two exponentials leads 
to the Pareto distribution\footnote{For the more realistic situation where
cities are on average growing, by an exponentially growing term $\epsilon_{i,t}$
so as to represent immigration or population fluxes across cities for instance,  the same reasoning applied
once a change of frame has been performed with respect to the exponentially
growing $\epsilon_{i,t}$ term (see \citeasnoun{sor98} for details).}.

\citeasnoun{Eeckhout2004}'s model provides an expression for the growth of cities
of the form (\ref{thj2tk21t}), with $a_{i,t} = 1/\Lambda^{-1}(1+ \sigma_{i,t})$ as defined on page 1447. 
The function $\Lambda(S_{i,t}) \sim S_{i,t}^\Theta$ denotes the net local size effect on the growth of cities, $\sigma_{i,t}$ corresponds to exogenous technology shock impacting city $i$ at time $t$ and 
$\Theta= -(\theta - \gamma - \beta/\alpha)$ in the notations of \citeasnoun{Eeckhout2004}. 
The exponents $\alpha$ and $\beta$ 
quantify the consumer preference with respect to consumption, amount of land and housing, and leisure.
The exponent $\theta$ describes the 
dependence of the positive externality of being in a city of size $S$.
The exponent $\gamma$ describes the dependence of the 
negative external effect of how leisure can be used for labor. 
Then, any mechanism, ensuring a minimum (even random) city size helping to transform
(\ref{thj2tk21t}) into (\ref{eq:motion}) or equivalent \cite{sor98}, leads to the Pareto distribution 
for the tail of the distribution of city sizes with tail exponent $\alpha=-\Theta$.
Since $\Theta <0$, the ``net local size effect" $\Lambda(S_{i,t})$ is an inverse power of the 
city size so that a faster decay of the tail of the distribution of city sizes corresponds to 
a weaker relative impact of net local externalities $\Lambda(S_i,t)$
on large cities compared to smaller cities.
Zipf's law is recovered for the special case $\Theta= -1$.

\section{Testing the Pareto against the lognormal distribution}

\subsection{The uniformly most powerful unbiased test}

As summarized in the introduction,  \citeasnoun{Eeckhout2004} and \citeasnoun{Levy2007} have used general tests (L-test and $\chi^2$-test respectively) of the null hypothesis that the whole sample or just the upper tail is generated by a lognormal distribution, and they reach opposite conclusions. While these two tests are quite versatile, they are not always very powerful. For the purpose of comparing the lognormal distribution with Zipf's law, their lack of power can be ascribed to the fact that they test the null hypothesis against any alternative distribution, and not specifically against the Pareto distribution. But the later is the alternative of interest.  For instance, figure 2 in \cite{Eeckhout2008} illustrates the dramatic lack of power of the L-test in the upper tail of the distribution under the null of a lognormal: the confidence bands derived from this test fan out very strongly, which makes this test completely unable to decide if the deviations observed in the data are genuine or fake. Of course, the main reason for the decreasing power observed in figure 2  in \cite{Eeckhout2008} is the shrinking sample size for the upper ranks, but this does not remove the necessity of using the most possible powerful test in such a situation.

The discussion following equations (\ref{thertbww}) and (\ref{tbwrtbqga}) suggests that it might be possible to clearly distinguish between the explanatory power offered by a lognormal distribution versus a Pareto distribution for the U.S. Census 2000 data sample, when using a more powerful test.
The most general test that addresses the core question, whether the Pareto law holds in the tail or the lognormal model is sufficient, is to consider the two hypotheses: Pareto distribution for values of $x$ larger than some threshold $u$ and lognormal distribution also for value of $x$ above the same threshold $u$. Specifically, 
we propose to test the null hypothesis that, beyond some threshold $u$, the upper tail of the size distribution of cities is Pareto
\be
\label{eq:Pareto}
H_0: \quad f_0(x;\alpha) = \alpha \cdot \frac{u^\alpha}{x^{\alpha+1}} \cdot 1_{x \ge u}, \quad \alpha >0,
\ee
against the alternative that it is a (truncated) lognormal
\be
H_1: \quad f_1(x;\alpha,\beta) = \[\sqrt{\frac{\pi}{\beta}} e^{\frac{\alpha^2}{4\beta}} \(1- \Phi \(\frac{\alpha}{\sqrt{2 \beta}}\)\)\]^{-1}\frac{1}{x}\, e^{-\alpha \ln \frac{x}{u} - \beta \ln^2 \frac{x}{u}} \cdot 1_{x \ge u}, \quad \alpha \in {\mathbb R},~ \beta >0,
\ee
where $\Phi(\cdot)$ denotes the CDF of the normal distribution.

This is equivalent to testing the null hypothesis that the upper tail of the log-size distribution of cities is exponential against the alternative that it is a (truncated) normal. For this later problem, \citeasnoun{DCP99} have shown that the clipped sample coefficient of variation ${\hat c} = {\rm min}(1, c)$ provides the uniformly most powerful unbiased test, where $c$ is the sample coefficient of variation defined as the ratio of the sample standard deviation to the sample mean. The critical point of the test can be derived with extremely high accuracy (even for very small samples) by a saddle point approximation \cite{DCP99,GJ2002} or by Monte Carlo methods. 

\subsection{Results}

The upper panel of figure~\ref{fig1} depicts the $p$-value of the test as a function of the lower threshold $u$ expressed in terms of the rank of city sizes represented in a logarithmic scale. The $p$-values have been calculated using the saddle point approximation \cite{DCP99,GJ2002}. Extensive Monte-Carlo simulations reproduce basically the same results.
Figure~\ref{fig1} indubitably shows that the size distribution of the 1000 largest cities or so, which include more than half of the total population, is Pareto. This confirms and makes more precise the claim of \citeasnoun{Levy2007}. For larger ranks, the $p$-value becomes very small, qualifying the lognormal distribution as the better model for the set of smaller cities. This explains \citeasnoun{Eeckhout2004}'s results.

The lower panel of figure~\ref{fig1} depicts Hill's estimate $\hat {\alpha^{-1}}$ of the inverse of the tail index $\alpha$ of the Pareto distribution \eqref{eq:Pareto} again as a function of city rank. This estimator is the best unbiased estimator for the inverse of the tail index\footnote{It is not possible to get an unbiased estimate for $\alpha$.} \cite{Hill75}. For the U.S. census 2000 data (blue upper noisy curve), the inverse of the tail index is approximately constant and fluctuates around the value $0.7$  for ranks less than one thousand or so, confirming the validity of the Pareto model over this range. For ranks
larger than one thousand, the Hill's estimate $\hat {\alpha^{-1}}$ deviates rapidly, confirming a deviation from
the Pareto model for the set of smaller cities. 

In the lower panel of figure~\ref{fig1}, we also show Hill's estimate $\hat {\alpha^{-1}}$ for 
ten random samples drawn from a lognormal distribution with parameters $\mu=7.28$ and $\sigma=1.25$ (red curves). One can observe the absence of a plateau, and therefore no well-defined exponent, thus disqualifying the Pareto model. The increase of $\hat {\alpha^{-1}}$ with rank is the expected signature of the fact that the lognormal density is rapidly decreasing, i.e., it goes to zero faster than any power law, so that its effective tail index is equal to infinity and its inverse is vanishing. Therefore, for very low ranks (largest cities), Hill's estimator should converge to zero for data generated by a lognormal distribution. 

The contrast between the U.S. Census 2000 data and the samples drawn from a lognormal distribution with parameters $\mu=7.28$ and $\sigma=1.25$ is striking and provides additional evidence in favor of the Pareto distribution for the upper tail. This makes clear that the Pareto and lognormal models are distinguishable in their tail for the 
available U.S. Census 2000 data sample.

\subsection{Pareto model versus Zipf's law}

Now that we have established that the tail of the size distribution of cities is Pareto, we turn 
to the question of whether this Pareto law is Zipf's law, i.e., whether the exponent is $\alpha=1$. 

\begin{figure}
\centerline{\includegraphics[width=0.75\textwidth]{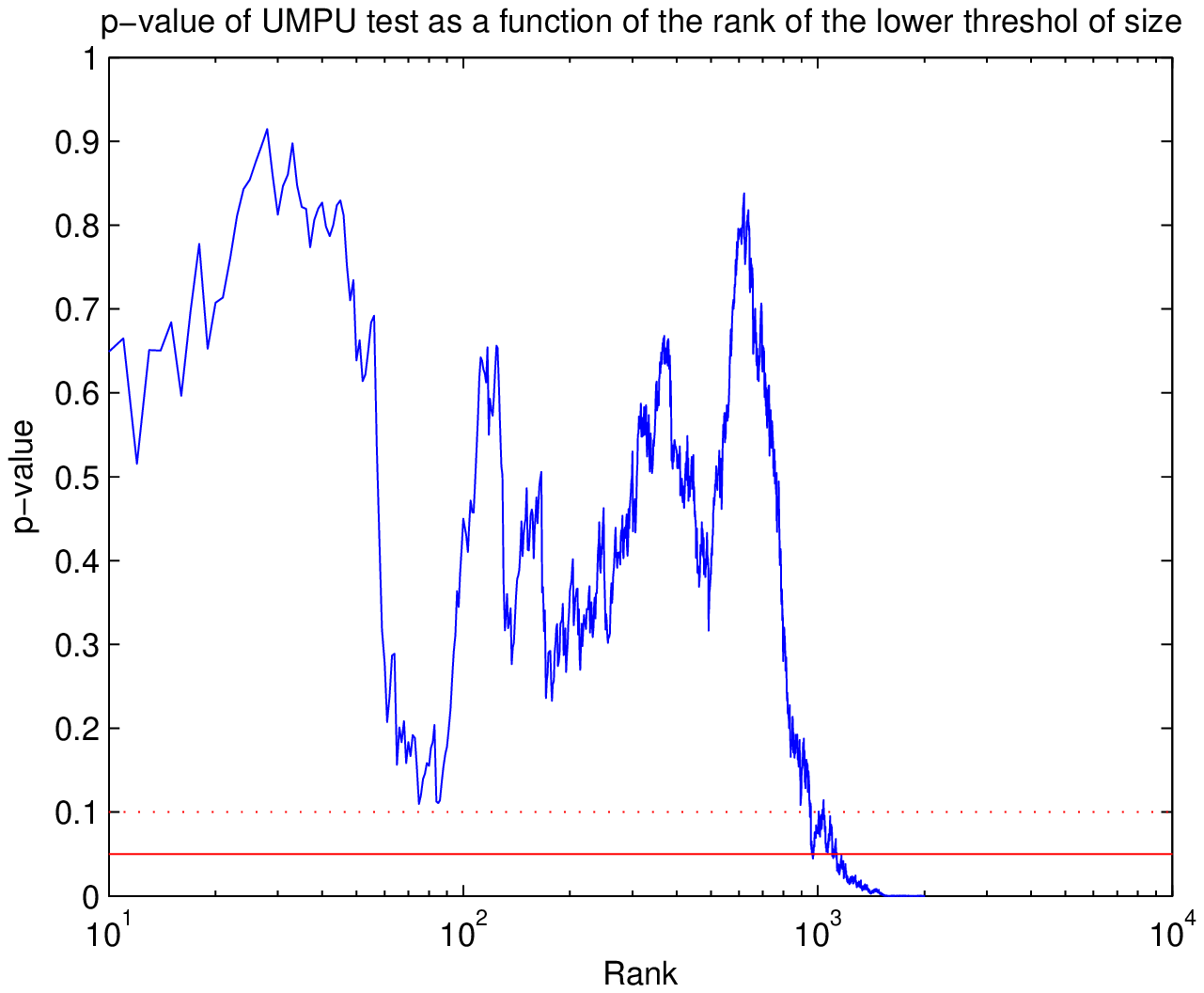}}
\centerline{\includegraphics[width=0.75\textwidth]{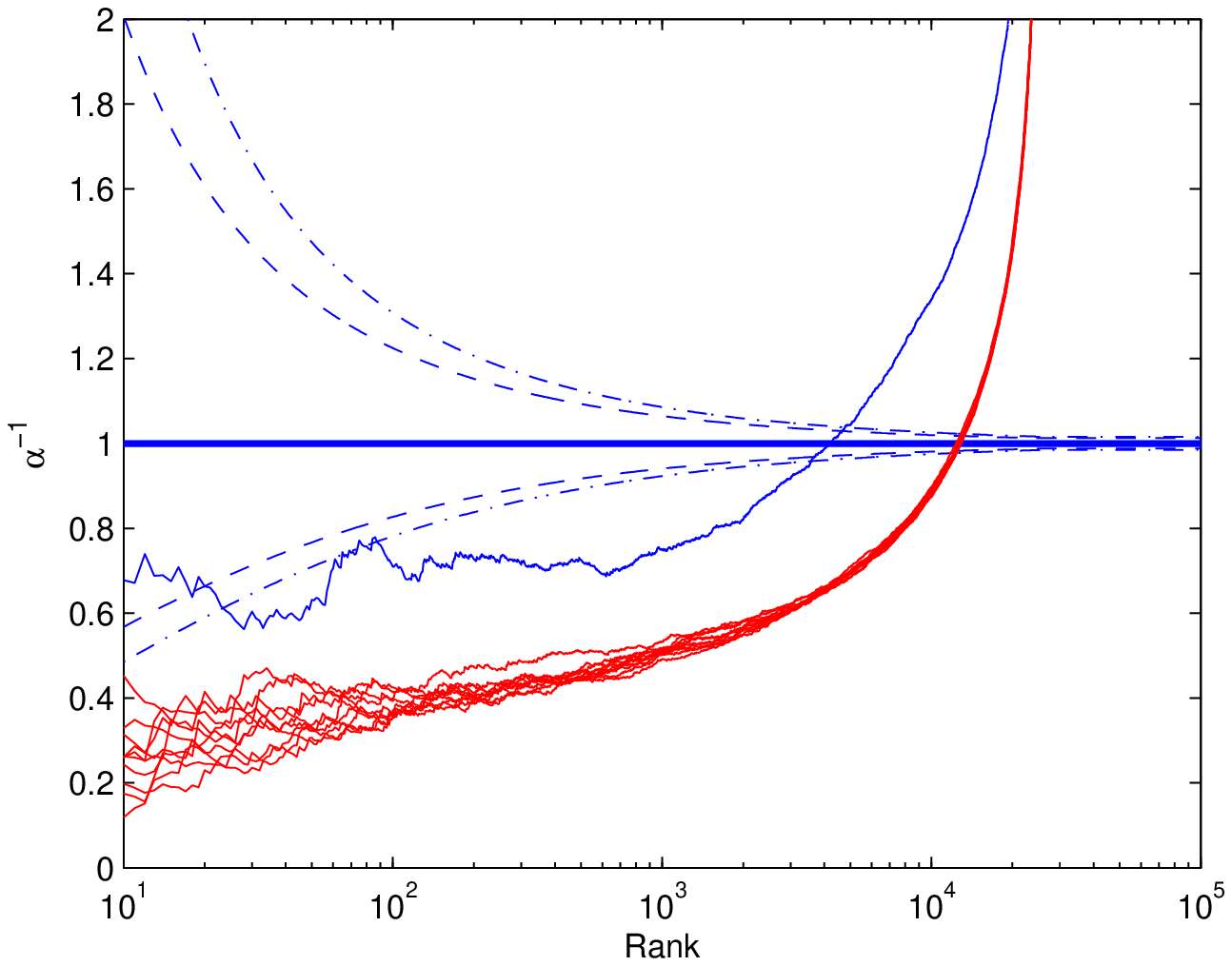}}
\caption{\label{fig1} The upper panel depicts the $p$-value of the test of the null hypothesis that the upper tail of the size distribution of cities is Pareto against the alternative that it is a (truncated) lognormal as a function of the rank threshold, where cities are ordered by decreasing sizes. The lower panel depicts Hill's estimate of the inverse of the tail index for the Census 2000 data (blue upper curve) and for ten samples drawn from a lognormal distribution with parameters $\mu=7.28$ and $\sigma=1.25$ (red bottom curves). The two dashed (respectively dot-dash) curves provides the confidence bands at the 5\%-significance level (respectively 1\% level) derived from the UMPU test that the tail index $\alpha=1$ against a two-sided alternative. }
\end{figure}

\begin{figure}
\centerline{\includegraphics[width=0.75\textwidth]{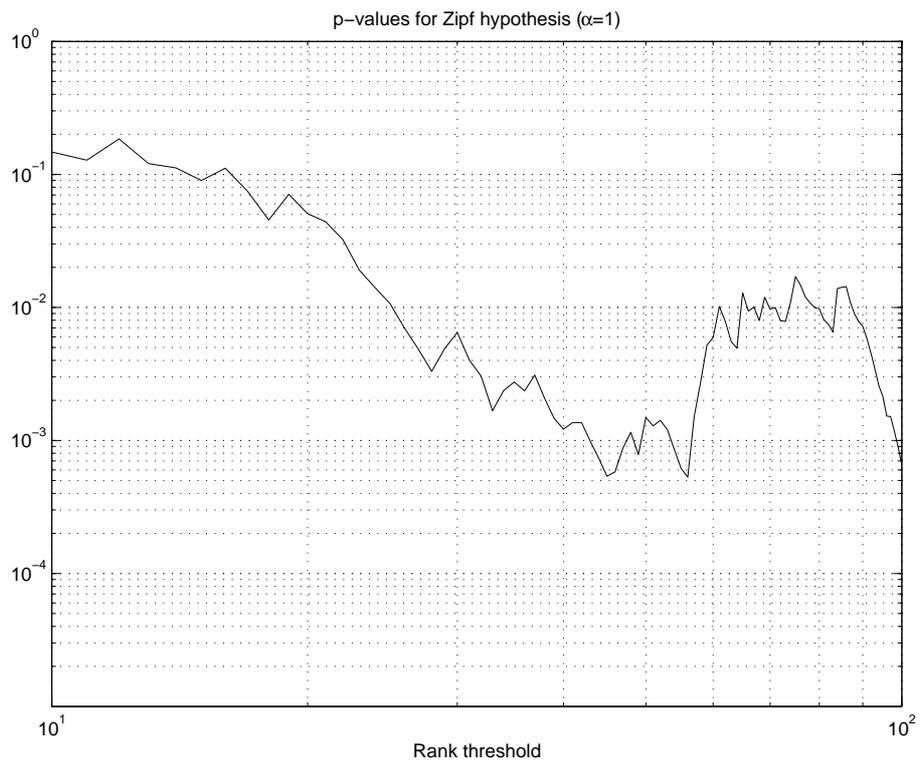}}
\caption{\label{fig2} One-sided $p$-value as a function of rank threshold, testing the hypothesis that the 
tail exponent of the Pareto distribution is compatible with Zipf's laws that $\alpha =1$. The $p$-value is
defined as the probability of exceeding the observed index estimate under the hypothesis that Zipf's law holds. }
\end{figure}

First, the lower panel of figure~\ref{fig1} shows the confidence band at the 95\%- and 99\% significance levels, derived from the uniformly most powerful unbiased test that the tail index $\alpha=1$ against a two-sided alternative \cite{LR2006}. At the 95\% significance level, Zipf's law is rejected, except for the twenty largest cities. 
Figure~\ref{fig2} improves on this statistics by plotting the $p$-value defined as the probability of {\it exceeding} the observed index estimate (one side-test) under the hypothesis that Zipf's law holds (index equals to unity). 
For rank thresholds larger than $20$, all $p$-values are smaller than $0.05$. For rank thresholds
larger than $16$, all $p$-values are smaller than $0.10$.
We are thus led to conclude that Zipf's law cannot be accepted
to describe the tail of the distribution of city sizes in the US census studied here, whereas a
larger exponent approximately equal to $1.4$ is significantly more likely. 

Coming back to \citeasnoun{Eeckhout2004}'s model, our finding $\alpha=-\Theta \approx 1.4$ implies
that the ``net local size effect" $\Lambda(S_{i,t})$ decreases faster with city size $S_{i,t}$ than would
be the case if Zipf's law held exactly. We also refer to 
\citeasnoun{SaichevetalZipf09} for a review of the mechanisms based on Gibrat's law leading to 
distributions with Pareto tails whose exponents can deviate from the Zipf's law value $\alpha =1$.

\clearpage

\end{document}